 \documentstyle[12pt]{article}
\begin{document}
 \baselineskip=24pt
 \parskip=0pt plus2pt
 \textheight=22cm

 \begin{titlepage}
 \begin{center}
{\LARGE\bf Structural Relaxation and Frequency Dependent Specific Heat in a\\
 Supercooled Liquid}
\end{center}
 \vspace{1in}
 \begin{center}
  Upendra Harbola and Shankar P. Das\\
 School of Physical Sciences, Jawaharlal Nehru University
 \\New Delhi 110067, India
 \end{center}

\vspace{1cm}
 \begin{center}
{\bf ABSTRACT}
\end{center}
\noindent
 We have studied the relation between the structural relaxation and the
 frequency dependent thermal                              
 response or the specific heat, $c_p(\omega)$, in a supercooled liquid.
 The Mode Coupling Theory (MCT) results are used to obtain
 $c_p(\omega)$ corresponding to different wavevectors. Due to the two-step
 relaxation process present in the MCT, an extra peak, in addition to the
 low frequency peak, is predicted in specific heat at high frequency.
\end{titlepage}
\section*{Introduction}
\noindent
 Understanding the complex relaxation behavior in supercooled liquids
 has been a field of much research interest in recent times.
 In this regard the response of a system to an energy fluctuation 
 namely the frequency dependence of specific heat of a supercooled
 liquid has been investigated by a number of authors.
 Generally, specific heat is a property that is usually linked to the 
 thermodynamic property of a system .
 The pioneering experiments done by Birge and Nagel 
 \cite {norman} have studied the dynamic response in glassy systems 
 namely glycerol and propylene and obtained interesting dynamical response 
 behavior, expressed in terms of a frequency dependent specific heat 
 \cite{march}. 
 In an experiment usually the frequency dependent product of thermal
 conductivity($\kappa$) and specific heat, $\kappa c_p(\omega)$ is
  measured. However, in the temperature range $190^oK-220^oK$, over which we 
 are interested here, $\kappa$ has weak frequency 
 dependence\cite{devid-pohl},\cite{DN}
 and hence the dynamics observed in the product $\kappa c_p(\omega)$
 is entirely due to the frequency dependence of the specific heat.
 The theoretical modeling for the  frequency dependence of specific heat 
 in a supercooled liquid has been studied by various authors 
  \cite {Oxtoby},\cite{Nielsen-Dyre}.
 New internal mode was proposed to be present in the supercooled liquid to 
 explain the frequency dependent response. In a simple analysis,
 Zwangig however had argued \cite{zwanzig}  that the frequency dependence of the 
 specific heat can be obtained without introducing any such internal mode.  
 This work showed that what is measured as the frequency dependent specific 
 heat  is actually related to that of the longitudinal viscosity in the 
 liquid. In this model the dynamics of fluctuations around the equilibrium
  was studied in terms of a simple set of slow variables of Hydrodynamics
 of fluids.
  These equations of motion used were the conservation laws of mass,
  momentum and energy in the system. The resulting formula for the
  specific heat is equivalent to
  linking of structural relaxation in a supercooled liquid to the 
 frequency dependence of the specific heat. 
 In the present work we take the data from the Specific heat measurement 
 \cite{norman} and extract the frequency dependence of the viscosity 
 as will be required from such a formulation proposed in Ref. \cite{zwanzig}. 
 We then address the question if this value of the longitudinal viscosity
 will indeed be self consistent with independent measurements on the
 structural properties. The basic idea is to consider the frequency
 dependence of the specific heat solely in terms of the structural
 relaxation in the supercooled liquid.

 For structural relaxation
- the microscopic model for the liquid dynamics namely the Mode Coupling
  Theory (MCT) has been studied by a number of authors in recent
 years \cite{gtze-les,gene-adv}. 
 A simple application \cite{march} of the MCT to fit the specific heat
 data indicated a very large exponent is required
 to match the experimental data to power law divergence.
 In the present work we also use the MCT as a model for structural 
 relaxation and obtain the 
corresponding frequency dependence of the longitudinal viscosity. 
 We then use it to predict the behavior of the
 specific heat with frequency as predicted from the theory proposed
 in Ref \cite{zwanzig}.
 In the microscopic model of the Mode coupling Theory the
  wave number dependence of  the longitudinal
 viscosity is obtained using proper input for the structure
 factor of the liquid.
 For this purpose standard results from the integral equations for
 simple liquids are used for the structure factor.
 The frequency dependent specific heat is then
 computed  for different wave numbers.
 Thus the effect of the response
 to heat fluctuations can be computed over different length and time
 scales in the present approach.
While this extends the theory with scope of further comparison,
 the main goal of the present work is to test if the frequency
 dependence of the specific heat can be understood solely in terms
 of the structural relaxation and if the two sets of measurements 
 agree in a self-consistent manner.
 The paper is organized as follows.
 In the next section, Sec. II, we consider the schematic model for the
 time dependence of the viscosity and, in  Sec. III, we compare the
 theoretical results with the experimental observations.
 A wavenumber dependent calculation for the specific heat is
 presented in the Sec.VI. In Sec. V, we present
 the Mode Coupling results for the specific heat
 over different length-scales and temperatures.
 In the last section we discuss the results.

\newpage
\section*{II. Frequency Dependent Specific Heat}
Since we are concerned here with the dynamic properties of a supercooled
liquid, an obvious choice is to consider a hydrodynamic model for the system.
To start with, we write down the linearized hydrodynamic equations,
\begin{equation}
\label{rho}
\frac{\partial}{\partial t}\delta\rho + \nabla.\vec{g}=0
\end{equation}
\begin{equation}
\label{velocity}
\frac{\partial  g_i}{\partial t}+\nabla_i P-\eta \nabla^2 \vec g_i
-({1\over3}\eta +\zeta)\nabla(\nabla.\vec g_i)=0
\end{equation}
\begin{equation}
\label{temp}
\rho_o c_v\frac{\partial}{\partial t}\delta T +\kappa \nabla^2 T+
 \frac{T_o}{\rho_o} 
\left(\frac{\partial P}{\partial T}\right)_{\rho} \nabla.\vec g_i=0,
\end{equation}
governing the time evolution of fluctuations of conserved variables- mass
density($\rho$), momentum density($\vec g$) and the temperature(energy)$T$.
Here $\rho_o$ and
$T_o$ represent equilibrium density and temperature respectively and
$\delta\rho$ and $\delta T$ are the fluctuations from the equilibrium values.
$c_v$ is the specific heat per unit mass at constant volume, $\eta$
and $\zeta$ are the shear and bulk viscosities respectively.
The viscosity coefficients here are divided by the density.
The fluctuation of the pressure $P$ around the equilibrium value
 can be expanded to the lowest order in density and 
temperature as,
\begin{equation}
\label{pressure}
\delta P=\left(\frac{\partial P}{\partial \rho}\right)_T \delta  \rho +
    \left(\frac{\partial P}{\partial T}\right)_{\rho} \delta T,
\end{equation}
\noindent
where we have assumed that the change of the Pressure functional
 with the density function at the equilibrium can be replaced by the
 equilibrium partial derivative - replaced by the corresponding 
 thermodynamic quantity.
Using the above equations the energy conservation equation 
(\ref{temp}) reduces to the Fourier heat law for thermal 
fluctuations,
\begin{equation}
\label{fourier}
\iota\omega\delta T=\mu(\omega)\nabla^2 \delta T,
\end{equation}

\noindent
 with the frequency dependent thermal diffusivity $\mu(\omega)$
 defined in terms of the specific heat $c_p$ as,
\begin{equation}
\label{defn-mu}
 \mu(\omega)= {{\kappa} \over  {\rho_o c_p(\omega)}} ~~.
\end{equation}
\noindent
The specific heat $c_p(\omega)$ is expressed in the form, 
\begin{equation}
\label{specific}
c_p(\omega)=c_v+(c_p-c_v){K_T(0)\over K_T(\omega)}.
\end{equation}
\noindent
The quantity $K_T(\omega)$ is called the generalized bulk modulus 
and is given by
\begin{equation}
\label{KT}
{{K_T(\omega)} \over {K_0}}
= 1 + \iota\omega\Gamma(\omega), 
\end{equation}
\noindent
 and is expressed in terms of the
 reduced form $\Gamma (\omega) = \eta_l(\omega)/c_o^2$ of the frequency
 dependent longitudinal viscosity $\eta_l(\omega)$.
In equation (\ref{KT}), $K_0$ is the $\omega =0$ limit
of $K_T(\omega)$.
Obviously for the liquid state with
 finite viscosity the zero frequency limit of $K_T(\omega)$ 
relates to the thermodynamic property of the supercooled liquid.
 The sound speed $c_o$ is given by 
\begin{equation}
\label{s-speed}
{c_o}^2 = \left( {\delta P \over \delta\rho}\right)_T 
= {K_0 \over {\rho_o}} ~~~.
\end{equation}

\noindent

A frequency dependent longitudinal modulus 
$M(\omega)$, the inverse of compliance,
 is defined along a similar line as,
\begin{equation}
\label{modulus}
{{M(\omega)} \over {K_0}} =\gamma + \iota\omega\Gamma(\omega),
\end{equation}
where $\gamma$ is the ratio of the long time limit of
 the specific heat $c_p(\omega=0)$ to $c_v$.
Equation (\ref{specific}) is the key formula used in this paper
for testing the idea of modeling the frequency dependence in
the specific heat solely in terms of the structural relaxation.
In obtaining equation (\ref{fourier}) one also needs to assume that
the following self-consistent relation holds,

\begin{equation}
\label{small}
\Delta (\omega) ~ \equiv ~
(\gamma -1){\omega 
\over
{\bar{M} (\omega) [ \omega + i \nu \bar{M} (\omega) ]}}
~~~ <<  ~~~ 1 ~~.
\end{equation}
\noindent
Here  we have expressed $M$ in the dimensionless form as,
$\bar{M} (\omega) = M(\omega)/K_0$. 
$\nu = c_0^2/\mu_o$, 
with $\mu_o = \kappa/(\rho_o c_v)$ is the bare thermal diffusivity.
We test the validity of the assumption (\ref{small})
in the frequency range where the analysis with respect to
 experimental data is made. 

\newpage
\section*{III. Comparison of Experimental Data}  
In this section we test the self-consistency in expressing
 the frequency dependent data on specific heat and Structural relaxation.
For the supercooled liquids the relaxation over longest time
 scales, i.e. the  $\alpha$-relaxations, follows the stretched exponential
 behavior, 
\begin{equation}
\label{KWW}
\eta(t)=\eta_o exp\left[-\left({t\over \tau}\right)^{\beta}\right]
\end{equation}
where $\eta_o$ is the amplitude and $\beta$ is the stretching parameter
which defines the degree of deviation from the exponential decay.
In fitting the specific heat data we use the dimensionless form for
 the specific heat ratio,
\begin{equation}
\label{spec}
c_p(\omega)=c_v \left [ 1 +
{{(\gamma -1 )} \over {1+\iota \omega \Gamma(\omega)}} \right ] ~~,
\end{equation}
\noindent
which reduces the formula in a dimensionless form. We fit the
specific heat data  
 of Ref. \cite{norman} to the formula (\ref{spec})
 using a simple stretched exponential (\ref{KWW}) 
relaxation function.
In Fig. 1(a) and 1(b) we show the respective fitting of the
 experimental data  for the real and
imaginary parts of $c_p(\omega)$ in the supercooled glycerol
for three different temperatures, T = 201.4$^oK$, 203.9$^oK$, 211.4$^oK$.
The arrows in Fig. 1(b) indicate the peak positions
in the imaginary parts of the viscosities at the corresponding 
temperatures.
In calculating the specific heat, the three parameters 
$\Gamma (t=0))$, the relaxation time $\tau$ and the
 stretching exponent $\beta$ are used as the free
parameters. $\gamma$=1.86.
Using the best fit values of the parameters with the specific
 heat data we compute the
structural properties of the liquid given by
 Modulus $M(\omega)$ defined in eqn. (\ref{modulus}) 
and the longitudinal viscosity. The resulting behavior
 for these quantities are compared with the experimental results
 as shown in Figures 2 and 3.

In Figure 2, we show the viscosity $\eta$ in the zero frequency 
 limit in units of $K_o\tau_o$ where $\tau_o$ is the unit of
 time used. In the Inset we show the corresponding experimental 
 data \cite{Angell,kivelson} for the viscosity. The experimental data
 shown here is over a much wider temperature range (317$^oK \sim 190^oK$)
 - both theoretical and experimental data agree with the
 Vogel Fulcher fit ( $\eta \sim exp[A/(T-T_o)]$ 
 for $T_o = 128^o$K and $A = 2480^o$K shown in both the figure
 and the inset as solid lines.
 The zero frequency modulus $K_o$ is roughly temperature independent
 \cite{letovitz}  over the range considered here.
 We notice that the viscosity increases by four orders of magnitude as
 the temperature is decreased over a small range($200^oK-220^oK$) near the
 glass transition temperature($T_g=$190$^oK$).
 We define a normalized longitudinal modulus 
 \begin{equation}
 \label{tildeM}
 \tilde{M} (\omega)={M(\omega)-M(0) \over M(\infty)-M(0)} ~~.
 \end{equation}
 \noindent
 In Fig. 3 (a) and (b) we show the real and imaginary parts
 of $\tilde{M}$ respectively denoted by $M^\prime(\omega)$
 and imaginary $M^{\prime\prime}(\omega)$  against the frequency
 for the three different temperatures, T=203.9$^oK$(solid), 211.4$^oK$
 (dashed), 221.5$^oK$ (dotted) lines.
 The frequency in each case is expressed in terms of the ratio with the
 corresponding peak position $(\omega_p)$ in the imaginary part.
 The corresponding results from measurements on the modulus $\tilde{M}$
 \cite{jeong-et-al} are also shown with filled circle.
 In Fig. 4 we show the plot of the peak positions as found in the
 fitting with different temperatures. 
 As the temperature is decreased, peak in the imaginary part of the 
 specific heat shifts towards the lower temperatures, signifying the
 slower relaxations in the system.
 The solid line indicate V-F fit with $T_o = 128^o$K. 
 In Fig 5 we show the frequency dependent Specific heat and
 the viscosity function ( in the inset ) at the same temperature. 
 The peaks appear nearly at the same position on the
 frequency scale for the two quantities showing that the 
 dominant time scales are same in the two cases.
Finally we test the validity
 of the assumption (\ref{small}) that is crucial in reaching the Fourier
 heat law (\ref{fourier}) with the frequency dependent specific heat - defined
 above.  
For this we calculate 
both the real$(\Delta^\prime(\omega))$ and imaginary
$(\Delta^{\prime\prime}(\omega))$ parts of $(\Delta(\omega))$ for the
supercooled glycerol. In Fig. 6 we plot both the real
and imaginary parts of $\Delta(\omega)$
on the frequency range over which specific heat(frequency dependent)
is observed. These figures clearly show that the quantity $\Delta(\omega)$
is much smaller as compared to unity over this frequency range. This
substantiates the assumption made in the previous section to reach
 the Fourier heat law in a generalized sense.
\section*{IV. Wavevector Dependence of Specific Heat}

 In the previous section, we studied the specific heat and other
 quantities like longitudinal viscosity and modulus using a schematic
 model to show the self-consistency of the relation between structural
 relaxation and the frequency dependence of specific heat.
 Here we consider the wavevector and frequency dependent specific heat
 in a liquid. Starting from the generalized hydrodynamic equations 
 for the conserved densities in q-space, we obtain an equation,
 \begin{equation}
 \label{qtemp}
 \iota\omega\rho_o c_v \delta T(q,\omega)=- q^2\kappa \delta T(q,\omega)
 +\frac{\iota q^2 \omega T_o}{(\rho_o\omega^2-q^2 K_T(q,\omega)}
 \left(\frac{\partial P}{\partial T}\right)^2_\rho \delta T(q,\omega)
 \end{equation}
  which describes dynamics of the energy fluctuations over different
  length and time scales. $K_T(q,\omega)$ is the wave vector and
  frequency dependent bulk modulus given by,
\begin{equation}
{K_T(q,\omega) \over K_T(q)}= 1+\iota \omega \Gamma(q,\omega)
\end{equation}
 where $\Gamma(q,\omega)$ is the wavevector and frequency dependent
 longitudinal viscosity devided by the square of the speed of
 sound $c^2_s(q)= K_T(q)/\rho_o$. $K_T(q)$
 is the zero frequency limit of $K_T(q,\omega)$.
 The energy equation (\ref{qtemp}) reduces to the wavevector
 dependent Fourier heat equation,
 \begin{equation}
 \label{qfourier}
 \iota \omega \delta T(q,\omega)= -q^2 \chi(q,\omega) \delta T(q,\omega)
 \end{equation}
 where $\chi(q,\omega)=\kappa/(\rho_o c_p(q,\omega))$ is thermal
 diffusivity and $c_p(q,\omega)$ is the $q$-dependent specific heat given by,
 \begin{equation}
 \label{qspecific}
 c_p(q,\omega)=c_v\left[ 1+(\gamma_q-1)\frac{1}{1+\iota \omega
 \Gamma(q,\omega)}\right]
 \end{equation}
 and $\gamma_q$ is the ratio $c_p(q)/c_v$.
 Here in obtaining the Fourier heat equation (\ref{qfourier}), we have
 assumed that the quantity,
  \begin{equation}
  \label{qgamma}
 (\gamma_q-1){\omega \over \bar{M}(q,\omega)[\omega+\iota\nu(q)
 \bar{M}(q,\omega)} ~~<< 1
 \end{equation}
 where $\nu(q)=c_s^2(q)/\mu_o$ and
 $\bar{M}(q,\omega)=\gamma(q)+\iota \omega \Gamma(q,\omega)$.
 is much smaller as compared to unity.  In the long wavelength limit
 this quantity reduces to $\Delta(\omega)$ given by Eq. (\ref{small}).
\section*{ V. Results from the Mode coupling Theory }

 In this section, we consider the time dependent longitudinal viscosity
 obtained from a microscopic theory of Statistical Mechanics,
 instead of taking inputs from experimental results as was done in
 the section III. We predict structural
 aspects i.e. the wave vector dependence in the specific heat
 at different frequencies.

 In the simplest form the self-consistent mode coupling theory 
 predicts a sharp transition of the supercooled liquid to 
 nonergodic phase. In later versions it was shown that due to 
 coupling of density fluctuations with currents this sharp
 transition is eliminated - the full model with the cutoff mechanism
 included is termed as the extended model. In the work we consider
  the extended model where the cutoff function is adjusted to
 obtain agreement with the viscosity of the supercooled Liquid
 to the results obtained from Computer simulations. The details
 of the model and the scheme for computation of the
 density correlation function using the proper cutoff function
 is presented elsewhere. We consider a one component
 Lennard-Jones system for computing the structural relaxation
 properties using the MCT. 
The temperature $T^*$ and density $\rho^*$ are expressed in
the standard units of $\epsilon/K_B$ and $\sigma^3$ respectively. 
$\epsilon$ is the unit of energy in a L-J system and
 $\sigma$ is the diameter of a particle.

In computing the dynamical behavior of the density
 correlation function we estimate the cutoff parameters of the theory
 so that the shear viscosity obtained from the self consistent
 MCT agrees with computer simulation results\cite{sudha}. For the simulation
 results on one component model we use the recent results of Rucco et. al.
 \cite{rucco} using special techniques that avoid the typical problem
 of crystallization in one component systems. From the self consistent
 results for the density correlation functions we compute the mode 
 coupling integrals for the longitudinal part of the memory function
 related to the decay of the density correlation functions.

 The longitudinal viscosity in the zero wave number limit is 
 shown in Fig. 7 for the temperature range around $T_c$.
  The longitudinal viscosity shown for the temperature range
 around $T_c/T$ less than 1,  follows the power law behavior.
The viscosity  diverges with exponent equal to 1.9 around $T_c$
and for lower temperatures
 ( $T_c/T~\rangle$ 1 ) the behavior follows a Vogel Fulcher form.
 This is usual with results of extended MCT \cite{utrecht}.
 We then use this
 frequency dependent memory function or the longitudinal viscosity
 to compute the corresponding frequency dependent specific heat.
 In the Mode-Coupling approximation, the normalized longitudinal
 viscosity, $\Gamma(q,\omega)$, is given by,
\begin{equation}
\label{mcgamma} 
\Gamma(q, \omega)= \frac{1}{2 n}S(q) \int \frac{d\vec k}{(2\pi)^3}
\left[\frac{\widehat q.\vec k}{q} c(k)+\frac{\widehat q.(\vec q-\vec k)}{q}
c(|\vec q-\vec k|)\right]^2 \psi(k,t) \psi(|\vec q-\vec k,t)
\end{equation}
where $\psi(q,t)$ and $S(q)$ represent density correlation function
and the structure factor respectively. $c(q)$ is the direct
correlation function of the system. $\widehat q$ denotes the unit vector
along $\vec q$ and $n$ is the number density.
Using the above expression for the longitudinal viscosity in Eq.
(\ref{qspecific}), we calculate the $q$-dependent specific heat in
 the supercooled liquid.

In figure 8(a) and (b) we show
 respectively the real and imaginary part of $c^*_p (q, \omega)$
 vs. the frequency.
This is shown here for three different wave vectors,
$q \sigma $=0, 7.05(peak of the structure factor) and 30 (upper cutoff
 taken for the k-integral) at temperature $T^*$=.559.
The inset of the corresponding figures shows the secondary peak predicted
 for fast processes at very high frequency window - representing
 the so called $\beta$-processes in the supercooled liquid.
The peak in the imaginary part shifts to lower frequency
 with lowering of temperature. In Fig. 9(a) we show the variation of the
 peak position with temperature of the liquid.
 The solid line in the figure shows Vogel-Fulcher fit with $T_o$=.014.
 In order to indicate the structural dependence we also
 show in Fig. 9(b),
 the dependence of the peak position on the wave number $q$
 at a fixed temperature $T^*$=.559 . The peak frequency signifying
 the dominant time scale for relaxation at different wavenumbers
 follows the nature of the structure factor.
 It shows a minimum at $q$-value which corresponds to the peak
 in the structure factor of the liquid.
 Successive minima in the figure correspond to the other less pronounced
 maxima in the structure factor.

\section*{ VI. Discussion}
 In section III, we considered two types
 of experimental measurements on the supercooled liquids, respectively
 related to the energy fluctuations and the structural relaxations.
 This was done to check the
 consistency of formula for specific heat obtained from simple
 analysis of the Hydrodynamic equations.
 The comparisons done in section III indicates that the frequency
 dependent specific heat can be understood in terms of the
 structural relaxation data in terms of the analysis proposed 
 in Ref. \cite{zwanzig}.

 Subsequently we apply the standard forms of the self-consistent
 Mode Coupling Theory to compute the frequency dependence of the
 viscosity and compute those for the specific heat. Here we use
  the formula in the terms of the Generalized Hydrodynamics, extending 
 the model to large $k$, or small wavelengths.
We find that the dispersion in the specific heat decreases
 as we go to the smaller length scales(higher $q^*$) with 
 corresponding increase in spectrum width. This 
 demonstrates the fact that at short length scales, the 
 relaxation is fast and here the memory effects reflecting cooperativity
 are not strong.

 The peak position ($\omega_p$) in the imaginary part of the 
 $c^*_p(q,\omega)$ shifts to higher frequency with the increase of
 the  corresponding wavevector $q^*$. 
 It however reaches to a minimum frequency at the structure factor peak.
 Finally since the MCT relates to the two step relaxation process
 in supercooled liquids, there is a corresponding implication on
 the specific heat curve predicting a peak at very high frequency in the
 specific heat. This is shown in the inset of figure 8(b).
 It is a consequence of secondary relaxation 
 in the supercooled liquid.
 Due to the constraints on the MCT at very low temperatures, we could
 not study the thermal response of the system close to the glass
 transition temperature $T_g$.
 However, as is shown in the Fig. 9(a), the main peak in the specific
 heat moves towards the smaller frequencies with decreasing temperature,
 thus at the temperatures very close to the glass transition one can
 expect the two peaks to lie further apart from each other.
 We have ignored here effects of nonlinearities
 in the energy equation \cite{gene-sph,zuberev}.
 This can produce frequency dependence on other transport coefficients
 like thermal conductivity  as well.
 However, observation of such a peak will further
  strengthen the validity of the simple analysis presented here in
  energy transport in terms of structural relaxation behavior.

\vspace{.5cm}
\section*{Acknowledgement}
We are greatful to Prof. N. O. Berge for providing data on specific heat
measurements. UH
acknowledges the financial support from the University Grant
Commission, India.

\vspace*{.5cm}
\noindent
\section*{Figure Captions}

 \vspace*{.5cm}
Fig. 1(a):
Theory fit to the experimental data(filled circles)
for the real part $c\prime_p(\omega)$ of the
specific heat ($c_p(\omega)$) at three temperatures $T$=
201.4$^oK$(continuous line), 203.9$^oK$(long dashed) and
211.4$^oK$(dotted). $\omega^*=\omega \tau_o$, where $\tau_o$
is the units of time used(see text).

Fig. 1(b):
imaginary part $c^{\prime\prime_p}(\omega)$ of
specific heat corresponding to the real
part shown in Fig. 1(a). arrows along the frequency
axis indicate the peak position in the imaginary part
of the corresponding viscosity.

Fig. 2:
temperature variation of viscosity in supercooled glycerol
 as obtained from the specific heat fitting. Viscosity is
given in units of $K_o\tau_o$.
experimental results for the viscosity
are shown in the inset. The
continuous line( both in the main figure and the inset) is the
Vogal-Fulcher, $\eta=\eta_oexp[A/(T-T_c)]$, fit
with $\eta_o$=1.08$\times 10^{-4}P$, $A$= 2480$^oK$ and
$T_o$=128$^oK$.

Fig. 3(a):
real part, $M^\prime(\omega)$, of the
normalized longitudinal modulus $\tilde M(\omega)$(see text) is plotted
at three different temperatures T=203.9$^oK$(solid line),
211.4$^oK$(dashed line) and 221.5$^oK$(dotted line). dots are the
experimental results of ref. \cite{jeong-et-al}.
frequency axis is scaled with respect to the peak values
$\omega_p$ for the three different temperatures.

Fig. 3(b):
Imaginary parts of the normalized longitudinal modulus
$\tilde M(\omega)$ corresponding to the real parts shown
in the figure 3(a).

Fig. 4:
 peak position $(\omega_p)$ in the imaginary part of the specific heat
 is plotted as a function of the temperature$(T)$. continuous line is the
VF fit :$\omega_p=\omega_o exp[-A/(T-T_c)]$; with $\omega_o=1.0
 \times 10^{15}$ Hz, $E$=2559.35$^oK$ and $T_o=$ 128.22$^oK$.

Fig. 5:
Imaginary part of the specific heat, $c^{\prime\prime}_p(\omega)$ at T=214$^oK$.
Arrow along the frequency axis at 2.45 indicates the peak position
in the imaginary part of the corresponding viscosity shown in the inset.

Fig. 6
real and imaginary parts of $\Delta(\omega)$(see text),
for supercooled glycerol, is plotted at $T=201.4K$.

Fig. 7:
mode-coupling viscosity is plotted as a function
of temperature. it shows a vogul-Fulcher behaviour,
$\eta_oexp[A/(T-T_o)]$
for $T<T_c$ with $T_o$=.023 while for higher
temperatures$(T>T_c)$ it follows a power low
behaviour with exponent 1.9 . Arrow along the temperature
axis at $T_C/T$=1.36 indicates the  power-low divergence.

Fig. 8(a):
MCT results for the real part of the normalized
specific heat $c^{*}_p(q,\omega)$
= $(c_p(q,\omega)-c_v)/(c_p(q)-c_v)$ for three wave vectors
$q*$=0 (dotted), 7.05(continuous) and 30(dashed) at
$T^*$=.559.  Here frequency $\omega^*$ is in the units of
the inverse of Lenard-Jones time,
$\tau=\sqrt{{m \sigma^2 \over \epsilon}}$.

Fig. 8(b):
imaginary part of $\tilde{c}_p(q,\omega)$ from the MCT
corresponding to the real parts shown in Fig 8(a).
Inset shows the secondary peaks predicted by the MCT
for the same three $q$-values.

Fig. 9(a):
variation of the peak position ($\omega_p$) in the
specific heat $\tilde {c}_p(q,\omega)$ with temperature
for $q$=0. along the y-axis, we have shown
$ \omega^*_p= \omega \tau \times 10^2 $.

Fig. 9(b):
variation of the peak frequency $(\omega_p)$ with wave vector
$q^*$. $\omega_p$ reaches to minimum at $q^*$=7.05 at which
the structure factor shows a maximum. $ \omega^*_p=
\omega \tau \times 10^2 $.

\end{document}